\documentclass[a4paper,11pt]{article}
\pdfoutput=1 

\usepackage{jinstpub} 
\usepackage{wrapfig}
\usepackage{float}
\usepackage{graphicx}

\title{\boldmath Light Yield and Uniformity Measurements of Different Scintillator Tiles with Silicon Photomultipliers}

\author[a,1]{G. Eigen,\note{Corresponding author.}}
\author[a]{and G. R. Lee}

\affiliation[a]{Department of Physics and Technology, University of Bergen,\\ Allegaten 55, 5007 Bergen Norway}

\emailAdd{gerald.eigen@uib.no}

\abstract{We present light yield and uniformity measurements of square and hexagonal tiles read out with silicon photomultipliers
via a Y11 wavelength-shifting fiber or directly from the side or from the center at the top face. All tiles are 3~mm thick and have an area of $\rm 9~cm^2$. The sides are wrapped with two layers of Teflon tape while top and bottom faces are covered with two layers of Tyvec paper. We further show the first light yield and uniformity measurements of ATLAS Tile Calorimeter (TileCal) tiles with MPPC readout. This study has been motivated by looking into a possible phase 3 upgrade for the ATLAS hadron calorimeter and for hadron calorimeters at future hadron colliders. 
}


\proceeding{Calorimetry for High-Energy Frontier- CHEF 2019\\
  25-29 November 2019\\
  Kyushu University, Fukuoka, Japan}

\begin{document}
\maketitle
\flushbottom

\section{Introduction}
\label{sec:intro}

We present herein light yield measurements of scintillator tiles read out by silicon photomultipliers (SiPM)\footnote{SiPMs  are called Multi-Pixel photon Counters (MPPCs) by Hamamatsu} in the context with  hadron calorimeter work for experiments at the ILC~\cite{adloff} and at hadron colliders~\cite{fcc},   
These include uniformity measurements of hexagonal tiles with different SiPMs and different photodetector couplings as well as ATLAS Tile Calorimeter (TileCal) tiles.  The idea of looking at the performance of hexagonal scintillator tiles arose by the SiD plans  to use hexagonal silicon pads in the electromagnetic calorimeter instead of square pads because the manufacturer obtains higher pad yields from a silicon wafer~\cite{sid}.
Using similar geometry cells in both the EM and hadron calorimeters improves the reconstruction of showers that start in the EM calorimeter and extend into the hadron calorimeter. Since a hexagon is a better approximation to a circle than a square we typically need to sum up  less tiles in the shower reconstruction yielding a better signal-to-noise ratio. This also improves the separability of two close-by showers. The study of  the ATLAS TileCal tiles with SiPM readout 
is of interest for a possible upgrade of the TileCal in phase 3 and for hadron calorimeters at future hadron colliders.

\begin{figure}[H]
\centering
\vskip -0.1cm
\includegraphics[width=95mm]{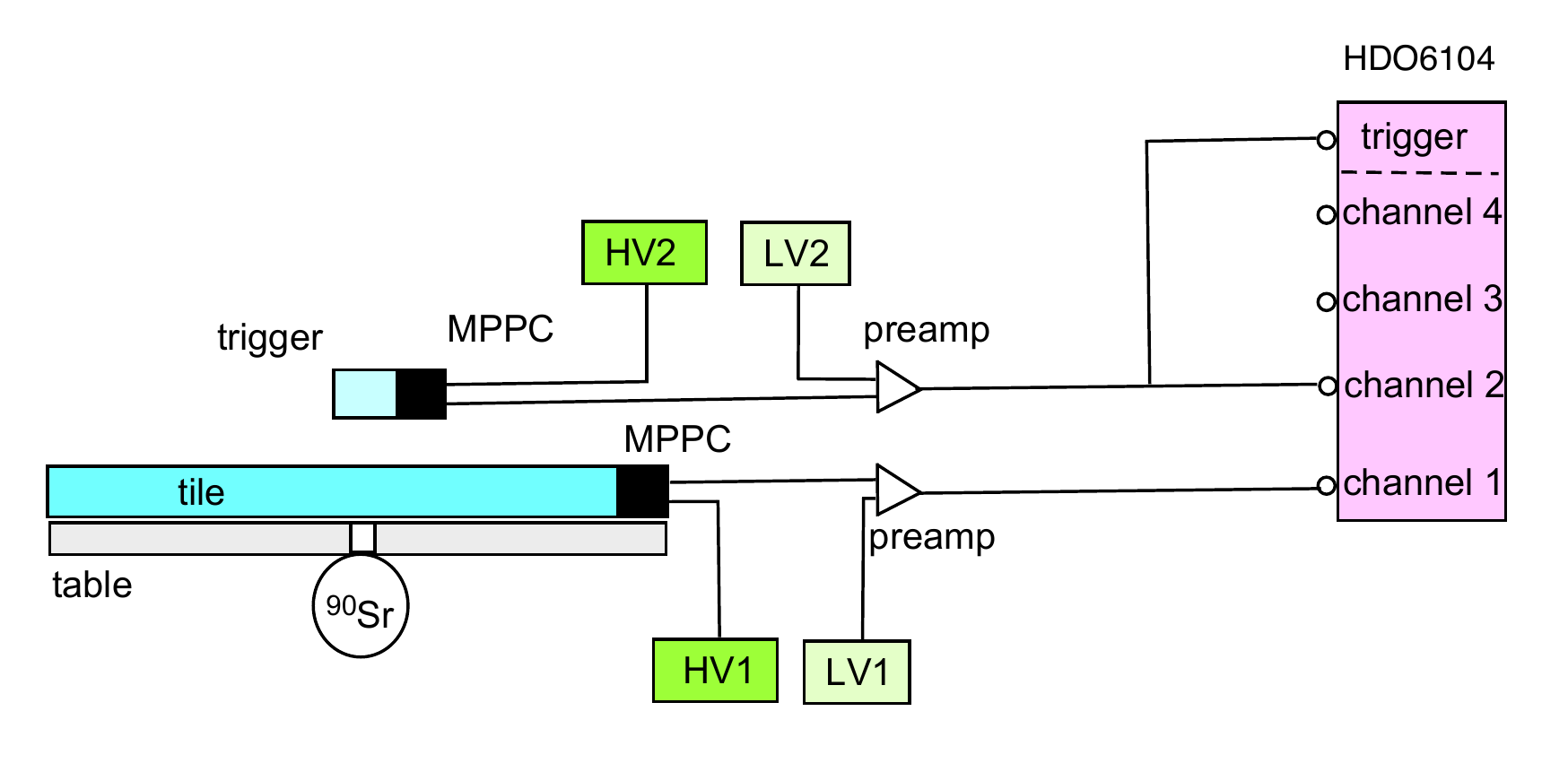}
\includegraphics[width=55mm]{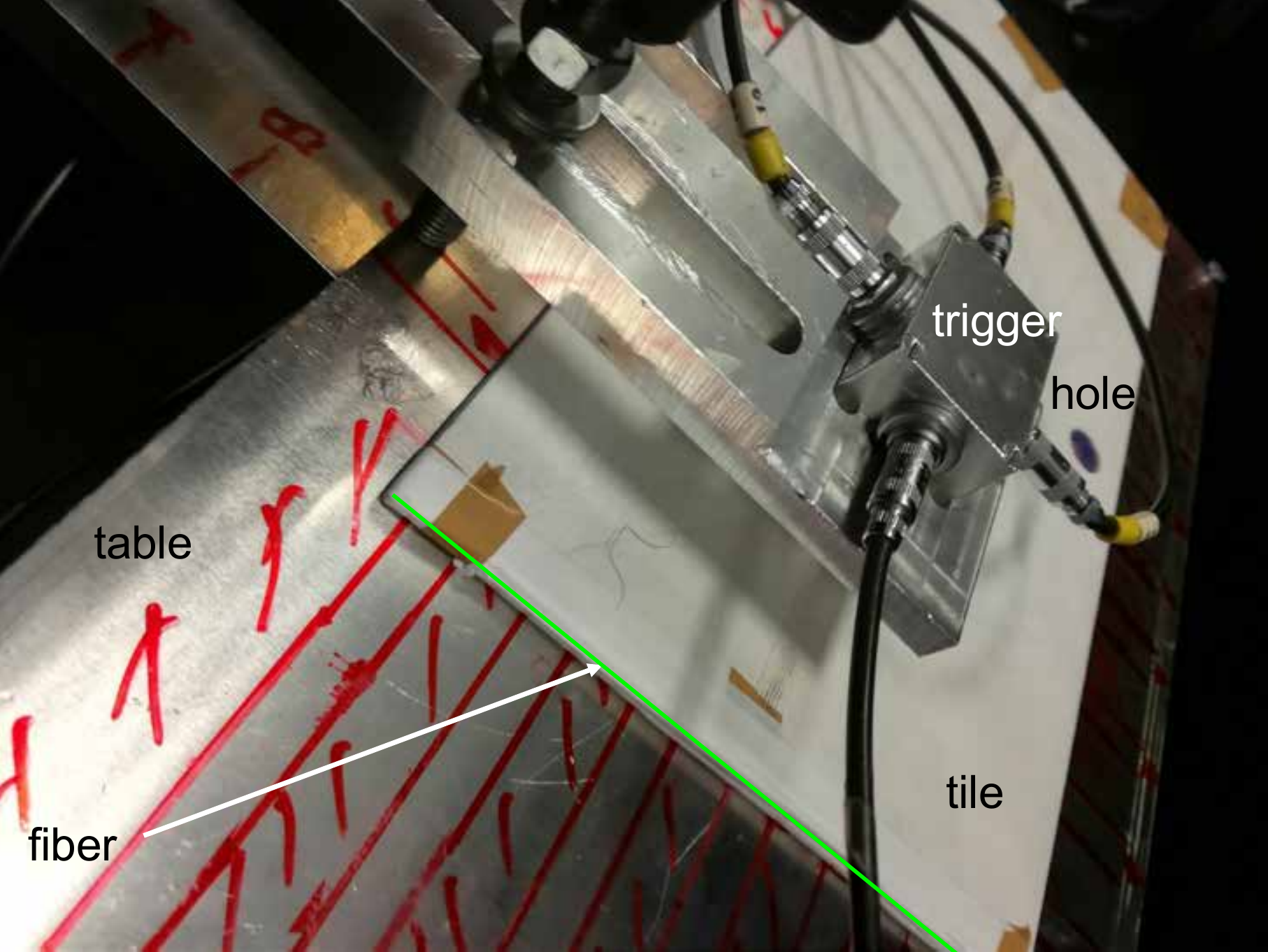}
\caption{Left: Schematic view of the measurement setup.  Right: Photograph of an ATLAS tile on the measurement table. 
}
\label{fig:setup}
\end{figure}

\section{Test of hexagonal and square  scintillator tiles with three different readout schemes}
\label{sec:tile}


Figure~\ref{fig:setup} (left)  shows a schematic view  of the measurement setup.  We perform all measurements inside a black box determining the light yield from the minimum-ionizing (MIP) peak of electrons from a  $\rm ^{90}Sr$ source~\cite{eigen}.
The tile under test is placed on a table with holes into which the $\rm ^{90}Sr$ source is inserted.  
The tile is read out with an MPPC, which is connected to a charge-sensitive preamplifier. Above the tile we place a second small trigger counter.  Whenever the trigger fires we record a waveform of the tile under test with a 12-bit digital oscilloscope. In each run we record 50 000 waveforms from which photoelectron spectra are extracted offline. Figure~\ref{fig:setup} (right)  shows a photograph of an ATLAS tile on the measurement table.

\begin{figure}[H]
\centering
\vskip -0.1cm
\includegraphics[width=60mm]{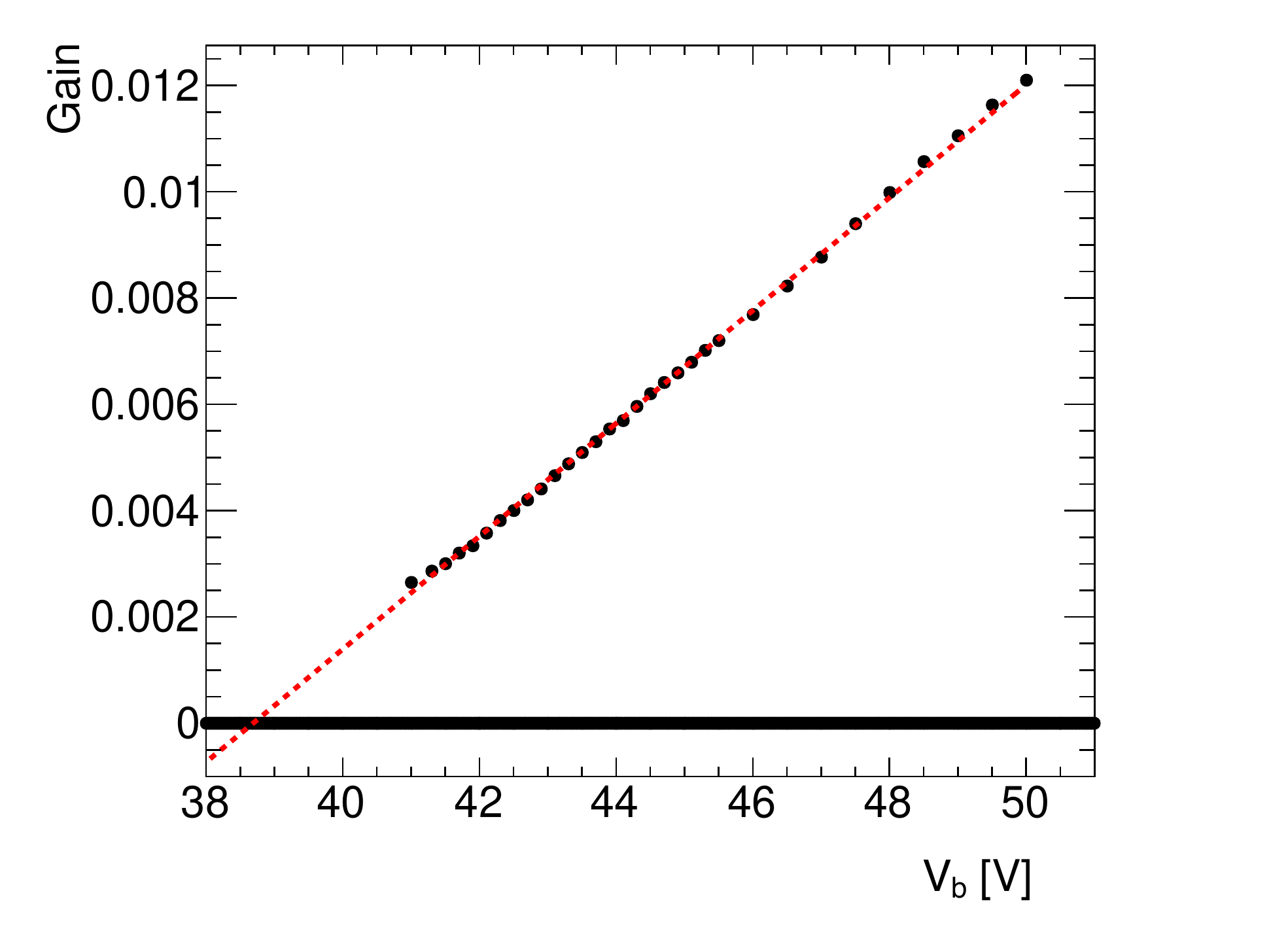}
\includegraphics[width=60mm]{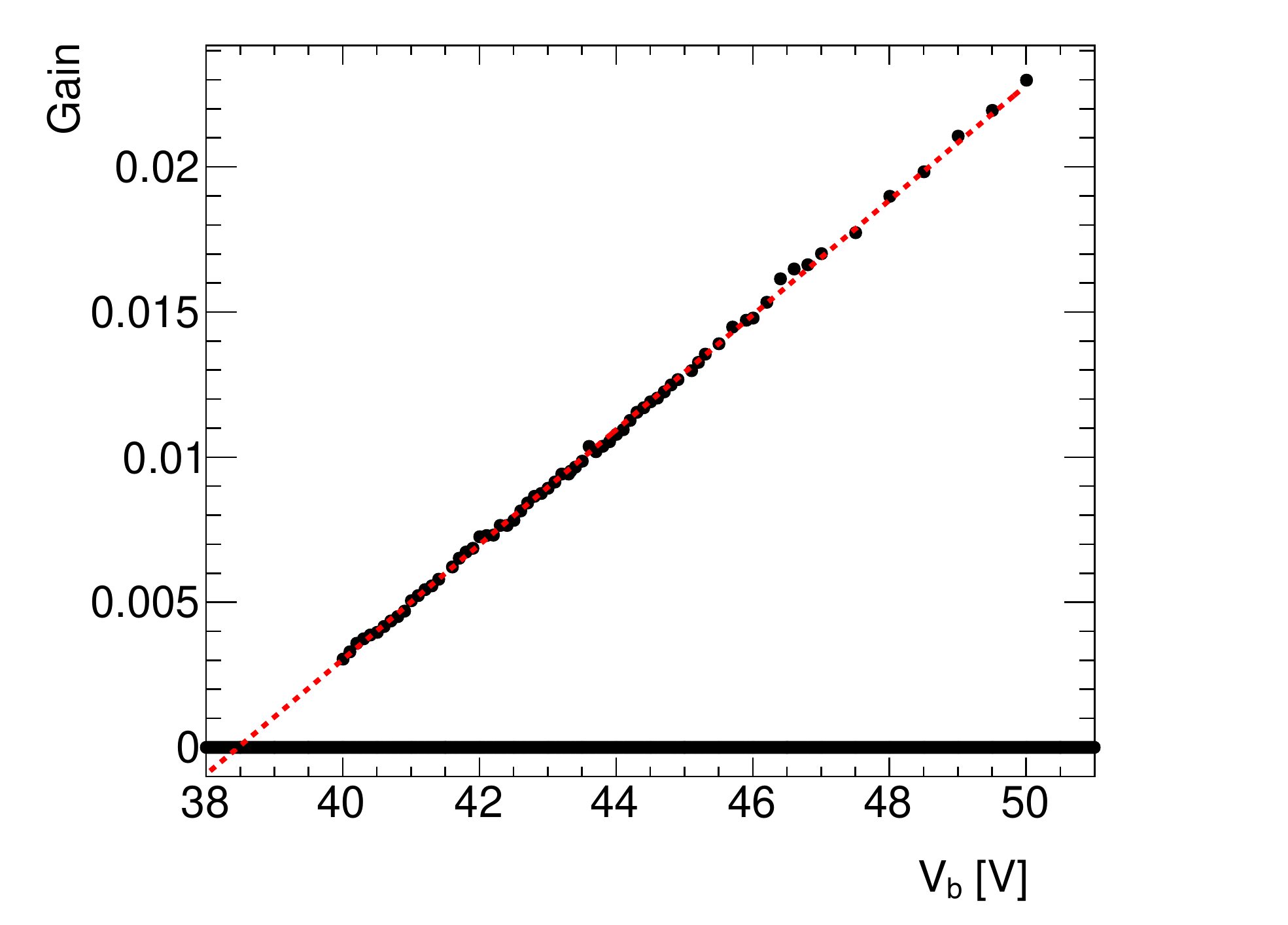}
\caption{Gain (arbitrary units) versus bias voltage for the 
S14160-1310  (left) and S14160-1315 (right).
}
\label{fig:gain}
\end{figure}

\begin{figure}[H]
\centering
\vskip -0.1cm
\includegraphics[width=107mm]{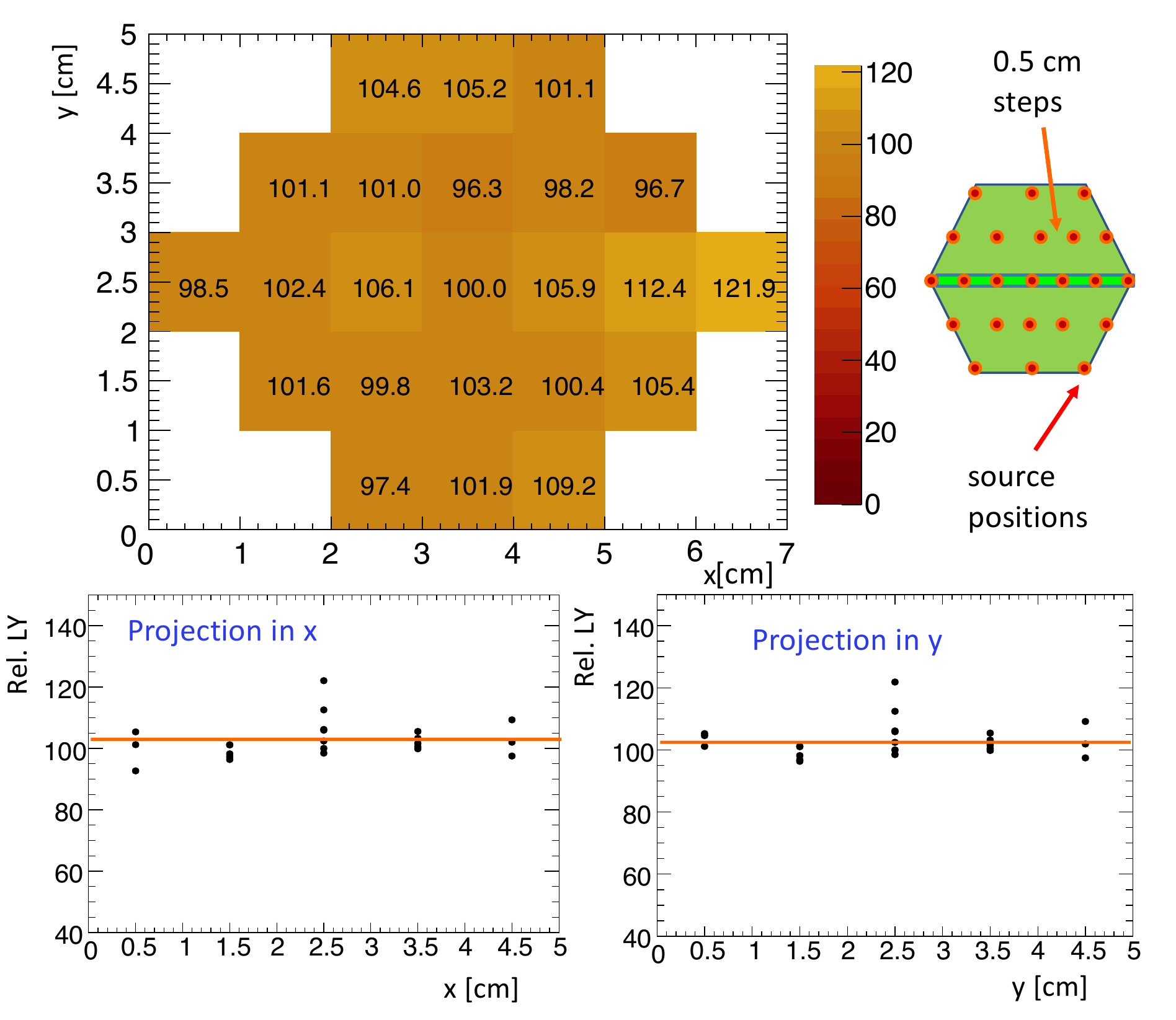}
\caption{Top left: Uniformity measurement of a hexagonal tile read out with a wavelength-shifting fiber. Top right: Source positions (red dots) and fiber position  (green line). Bottom left: Relative light yield in the $x$ direction. Bottom right: Relative light yield in the $y$ direction.
}
\label{fig:T1}
\end{figure}

We have  tested 3~mm thick  hexagonal and square tiles that  have an area of $\rm 9~cm^2$.
The scintillator material is BC404 from St. Gobain, which has an attenuation length of 140~cm and 
a wavelength of maximum emission at 408~nm. We use three different schemes to read out  the tile: i) via a green Y11 wavelength-shifting fiber from Kuraray that is inserted into a groove in the tile with  one fiber end attached to the MPPC and the other covered by a mirror; ii) placing the MPPC above a dimple in the center of the tile top face; iii) placing the MPPC on a tile side.
Top and bottom faces of the tiles are covered with two layers of Tyvec paper, while all sides are wrapped with two layers of Teflon tape. A 1~mm readout hole is punched into the wrapping. Outside the hole we placed  a  $3 \times 3 ~\rm mm^2$ MPPC (S13360-3025) or one of the new fourth-generation MPPCs (model S14160) that have $10~\rm \mu m$ or $15~\rm \mu m$ pixels. 
Figure~\ref{fig:gain} shows the gain versus reverse bias voltage dependence for the $1.3 \times 1.3~\rm mm^2$ MPPCs. The gain is linear over a range of 40-50~V.

\begin{figure}[H]
\centering
\vskip -0.1cm
\includegraphics[width=100mm]{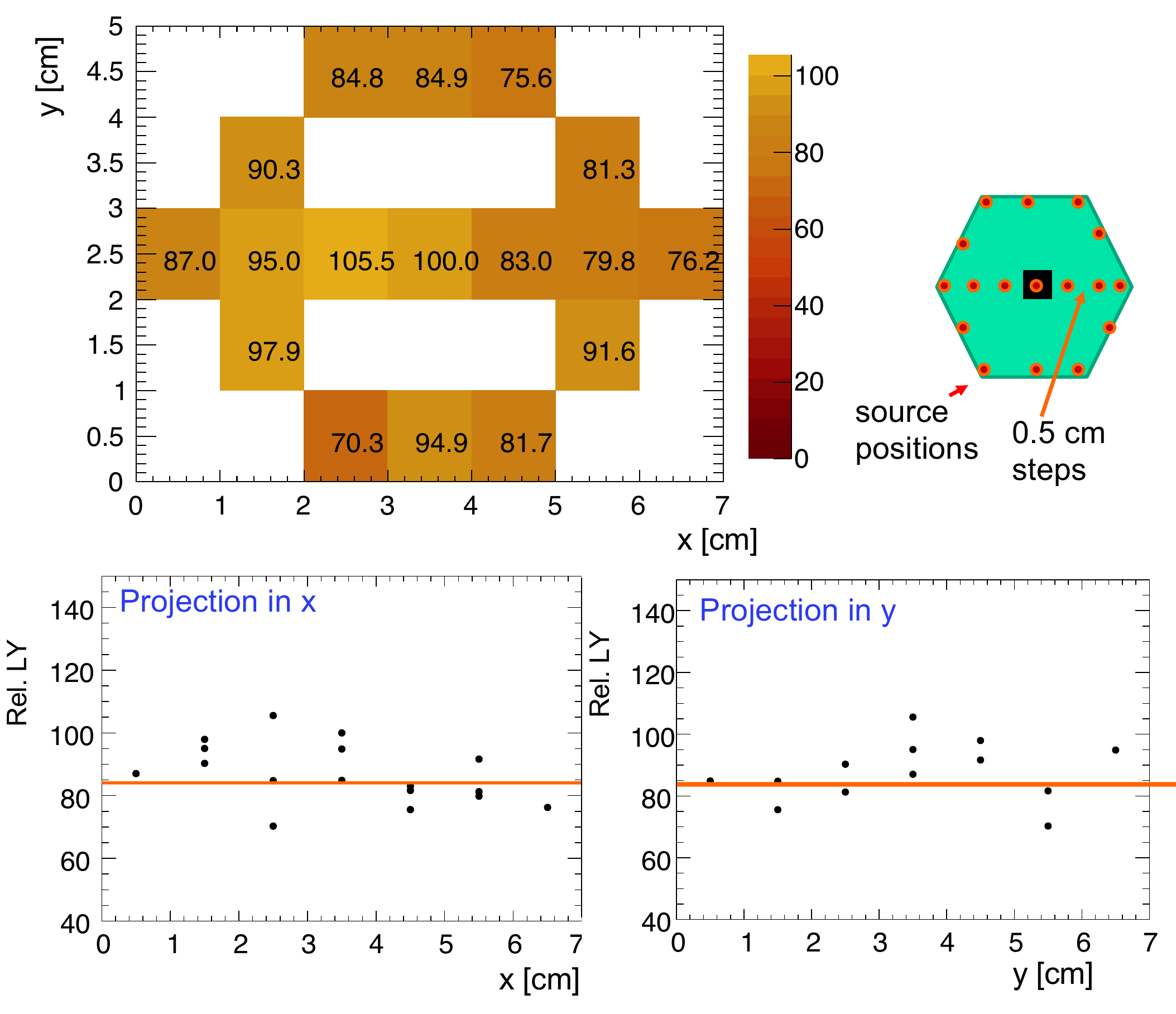}
\caption{Top left: Uniformity measurement of a hexagonal tile read out with SiPM at the top face center (black square). Top right: Source positions (red dots). 
Bottom left: Relative light yield in the $x$ direction. Bottom right: Relative light yield in the $y$ direction.
}
\label{fig:T2}
\end{figure}

\begin{figure}[H]
\centering
\vskip -0.1cm
\includegraphics[width=100mm]{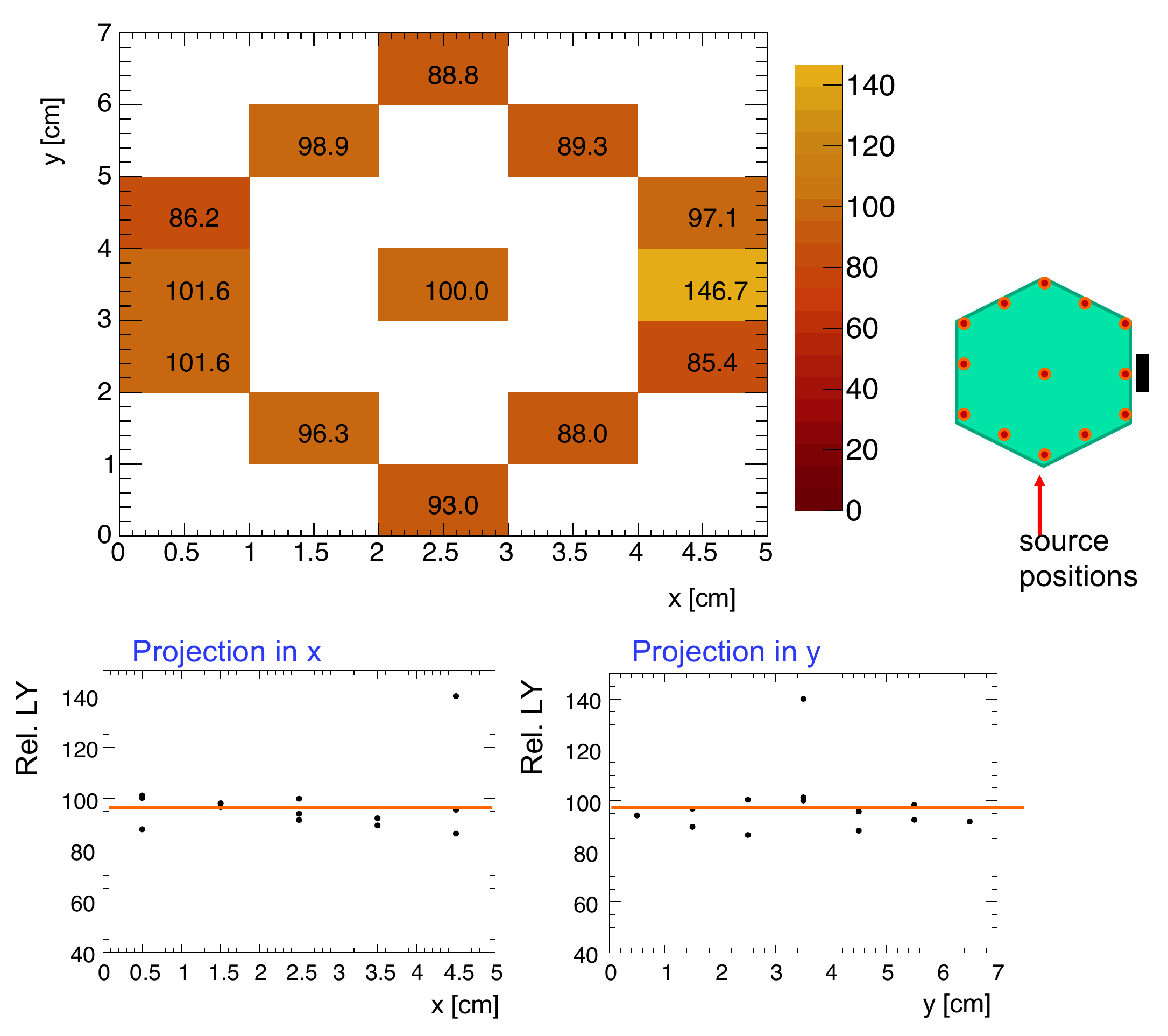}
\caption{
Top left: Uniformity measurement of a hexagonal tile read out with SiPM at the tile side (black square). Top right: Source positions (red dots). 
Bottom left: Relative light yield in the $x$ direction. Bottom right: Relative light yield in the $y$ direction.
}
\label{fig:T3}
\end{figure}

Figures~\ref{fig:T1},~\ref{fig:T2} and ~\ref{fig:T3} respectively show our relative light yield measurements of  hexagonal tiles read out with a fiber, with an MPPC placed on the center of the tile top face and an MPPC placed on the tile side. We have normalized the light yields to the value measured at the center position of the tile.
For the fiber readout the measurements  are rather uniform except for positions near the fiber end at the MPPC (right-hand side), where the light yield is a factor of 1.2 higher than that at the tile center.  
For the readout  at the tile center and the tile side we observe a rather uniform response except for positions near the MPPC where   light yields increased by factors of 1.2 and 1.48, respectively.  We obtain similar results fro the uniformity measurements of the square tiles.
To improve uniformity near the MPPC we will redo the measurements with a small MPPC encompassed in a sufficiently large dimple. Table~\ref{tab:ly} summarizes  light yield measurements at the tile center, relative light yields across  tiles and range of uniformity for hexagonal and square tiles for the three different readout schemes.  The low light yield of the square tile with readout at the tile top face center may be due to imperfect wrapping. To reduce the light yield spread across the tile we will redo the measurements  with tiles wrapped in 3M reflector foil.

\begin{table}[htbp]
\centering
\caption{\label{tab:ly} Light yield of the center tile and average relative light yield of hexagonal tiles and square tiles for three different readout schemes, via wavelength-shifting fiber, with MPPC on the tile side and with MPPC at the center of the top face. The uniformity range is defined such that 90\% of the measurements fall inside. Note that we always exclude the measurement near the MPPC.}
\smallskip
\begin{tabular}{|l|c|c|c||c|c|c|}

\hline
readout &light yield &  rel. light yield & uniformity & light yield &  rel. light yield & uniformity \\
  & hexagonal &hexagonal & hexagonal  & square & square & square  \\
\hline
fiber & 14.9 pe & $(102.2\pm1.0)\% $& $\pm 5\%$&  19.4 pe & $(85.4\pm 1.9)\% $ & $\pm 10\%$ \\
center & 11.8  pe& $(85.9 \pm 2.2)\%$  & $\pm 12\%$&4.0  pe & $(77.4\pm 3.9)\%$& $ \pm 9\% $  \\
side & 9.5  pe & $(93.9\pm 1.8)\%$& $ \pm 7.5\%$ &7.9  pe&$ (86.8\pm 3.2)\%$& $\pm 13\%$ \\
\hline
\end{tabular}
\end{table}


\section{Readout of ATLAS tiles with MPPCs}

\begin{figure}[h]
\centering
\vskip -0.1cm
\includegraphics[width=115mm]{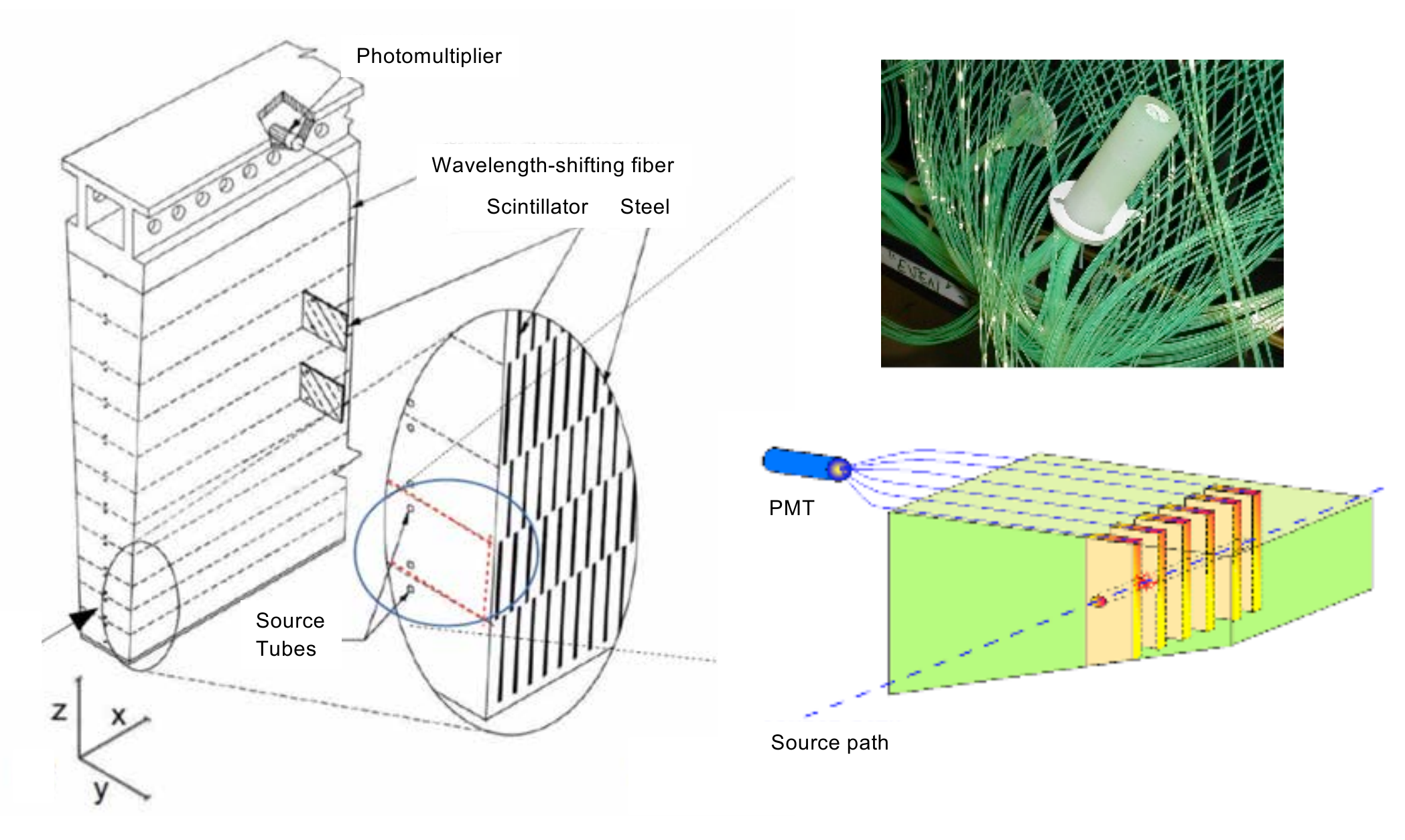}
\caption{Left: Schematic layout of the ATLAS TileCal module. Top right: Fibers bundled to couple to a photomultiplier. Bottom right: Fiber collection for the photomultiplier and path for the source calibration. 
}
\label{fig:tilecal}
\end{figure}

The ATLAS TileCal is a sandwich of scintillating tiles read out by wavelength-shifting fibers and PMTs and steel absorber plates~\cite{atlas}. Figure~\ref{fig:tilecal} left shows a schematic view of a calorimeter module. 
ATLAS uses three tile sizes, $12 \times 26~\rm cm^2$, $14.5 \times 30~\rm cm^2$ and $18.5 \times 35~\rm cm^2$. The tiles are slightly tapered and are read out via two Y11 wavelength-shifting fiber that are placed along the two tapered tile sides. As shown in Fig.~\ref{fig:tilecal} (top right) the fibers are collected into a bundle that is pushed into  a cylinder that in turn is coupled to  a photomultiplier tube via an air gap.  Some of the tiles have holes through which a $^{137}\rm Cs$ source is shot for calibrating the tiles. To measure the uniformity of the tiles, we built a table with a grid  of holes into which a $\rm ^{90}Sr$ source can be placed. Using only one fiber that was coupled to an MPPC, we measured the light yield of each tile at 15 positions (5 in $x$ direction that is perpendicular to the fiber and for 3 in $y$ direction that is parallel to the fiber) and normalized it to that at the center position. 

\begin{figure}[h]
\centering
\vskip -0.1cm
\includegraphics[width=90mm]{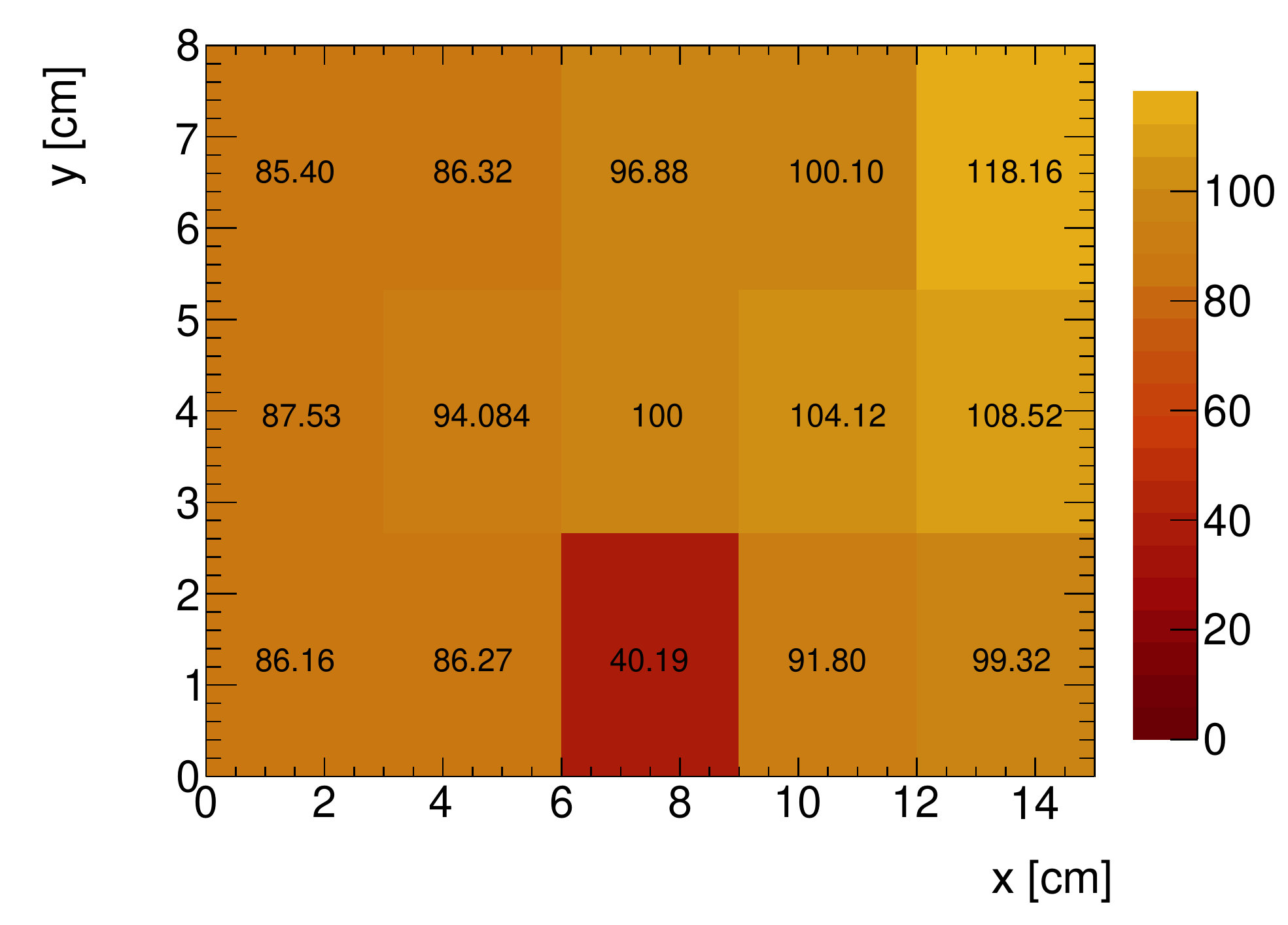}
\includegraphics[width=70mm]{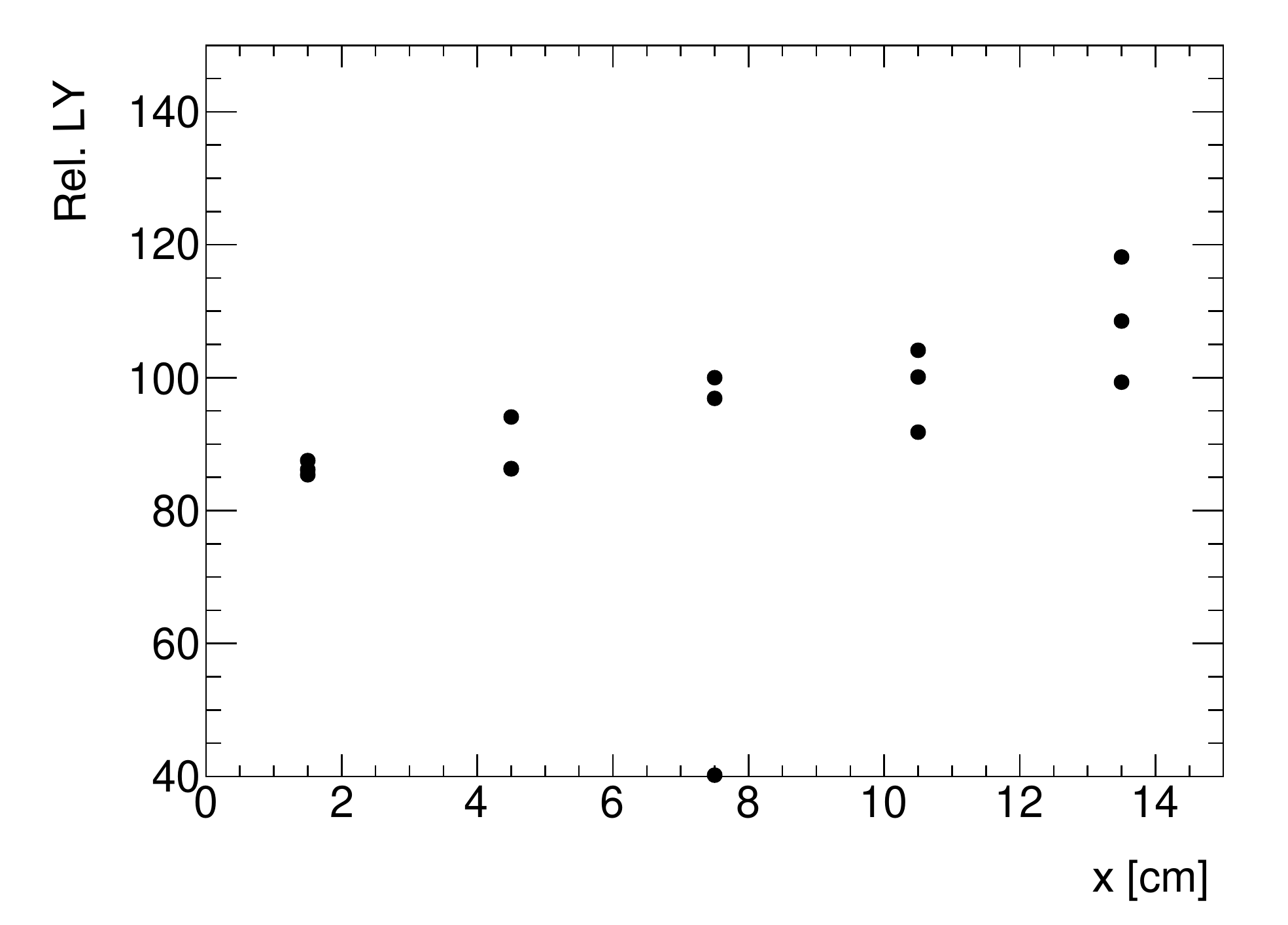}
\includegraphics[width=70mm]{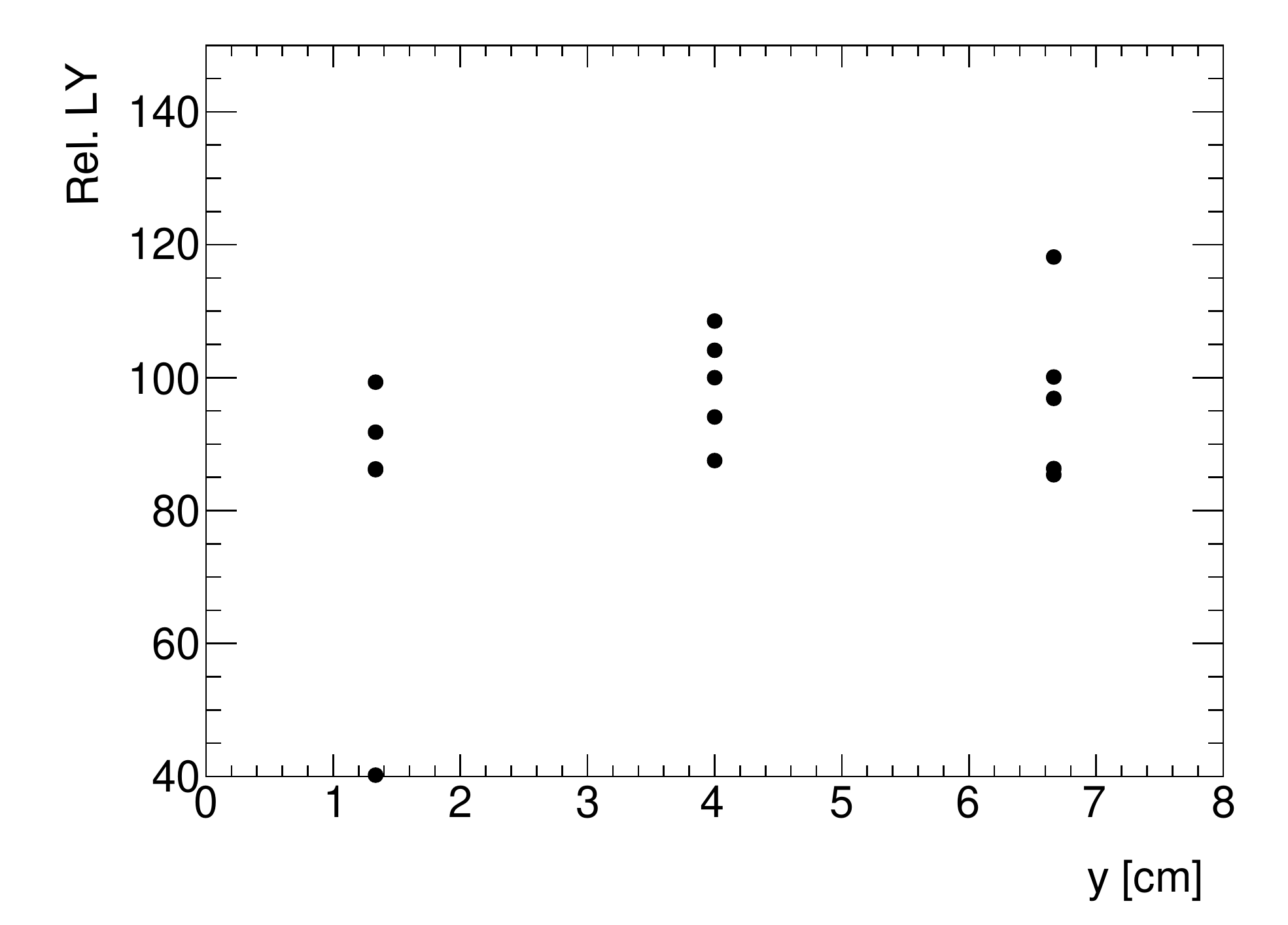}

\caption{Top: Uniformity measurement of a small ATLAS tile in the $x-y$ direction. The fiber is located on the right-hand side. Bottom: 
 Uniformity measurement of a small ATLAS tile in the  $x$ projection (left) and  $y$ projection (right).
}
\label{fig:tilecal-unif}
\end{figure}

After subtracting the pedestal, we fit the  photoelectron spectra with  Landau distributions to determine the MIP position. Figure~\ref{fig:tilecal-unif} (top) shows
a two-dimensional distribution of the relative light yield for a small-size ATLAS tile. Figures ~\ref{fig:tilecal-unif} (bottom) show the projections 
 in the direction perpendicular ($x$ projection) and parallel ($y$ direction) to the fiber, respectively. 
 We observe a slight increase in light yield in the $x$-direction, which is seen  for all other tiles as well. Uniformity is at a level of 15\%. 
The MIP peak lies around 3-4 photoelectrons, which is sufficiently large to read out all TileCal  tiles with MPPC arrays, 
 ensuring a large dynamic range. For tiles with calibration holes, the relative light yield near a hole is reduced to $\sim 40\%$.

\section{Conclusions and outlook}

The performance of hexagonal tiles looks very promising.  We will continue with these studies using enlarged dimples, different wrappings  and readout with the fourth-generation MPPCs. Furthermore, we need to measure more properties of the fourth-generation MPPCs including afterpulsing and dark current as a function of the bias voltage, the temperature dependence of the gain and the dependence of impinging photons to fired pixels (non-linearity). We have shown that ATLAS TileCal tiles can be read out with MPPCs. We will look at the readout with two fibers before reading out a fiber bundle with an MPPC array.


\acknowledgments

This work was supported by the Norwegian Research Council. We would like to thank Hamamatsu for supplying fourth-generation MPPCs.

\end{document}